\documentclass[format=sigconf,anonymous=false,authorversion=false]{acmart}

\usepackage[english]{babel}
\usepackage{booktabs}
\usepackage{moreverb}
\usepackage{multirow}
\usepackage{graphicx}
\usepackage[flushleft]{threeparttable}
\usepackage[utf8]{inputenc}
\usepackage{xspace}

\newcommand{\cfig}[4]
{
  \begin{figure}[#1]
    \centering
    \includegraphics[width=#2\columnwidth]{fig/#3}
    \caption{#4}
    \label{fig:#3}
  \end{figure}
}
\newcommand{\fig}[3]{\cfig{#1}{1}{#2}{#3}}

\newcommand{\fsource}[1]{\footnote{Source: #1}}
\newcommand{\furl}[1]{\fsource{#1}}

\newcommand{\figref}[1]{Figure~\ref{fig:#1}}

\newcommand{\tabref}[1]{Table~\ref{tab:#1}}
\newcommand{\secref}[1]{Section~\ref{sec:#1}}

\newcommand{\seefig}[1]{(see \figref{#1})}

\newcommand{\seetab}[1]{(see \tabref{#1})}

\newcommand{\daysForUsers}{6\xspace}
\newcommand{\monthsForTweets}{68\xspace}

\newcommand{\DataNamesNum}{19766\xspace}
\newcommand{\DataNamesNumFemale}{8351\xspace}
\newcommand{\DataNamesNumMale}{10467\xspace}
\newcommand{\DataNamesNumUnisex}{948\xspace}

\copyrightyear{2017}
\acmYear{2017}
\setcopyright{acmlicensed}
\acmConference{WebSci '17}{June 25-28, 2017}{Troy, NY, USA}
\acmPrice{15.00}
\acmDOI{http://dx.doi.org/10.1145/3091478.3091490}
\acmISBN{978-1-4503-4896-6/17/06}

\begin{document}

\title{Predicting Rising Follower Counts on Twitter Using Profile Information}

\author{Juergen Mueller}
\affiliation{
  \institution{University of Kassel}
  \department{Research Center for Information System Design (ITeG)}
  \streetaddress{Pfannkuchstr. 1}
  \city{34121 Kassel}
  \country{Germany}
}
\email{mueller@cs.uni-kassel.de}

\author{Gerd Stumme}
\affiliation{
  \institution{University of Kassel}
  \department{Research Center for Information System Design (ITeG)}
  \streetaddress{Pfannkuchstr. 1}
  \city{34121 Kassel}
  \country{Germany}
}
\email{stumme@cs.uni-kassel.de}

\begin{abstract}


When evaluating the cause of one's popularity on Twitter, one thing is considered to be the main driver: Many tweets. There is debate about the kind of tweet one should publish, but little beyond tweets.
Of particular interest is the information provided by each Twitter user's profile page. One of the features are the given names on those profiles. Studies on psychology and economics identified correlations of the first name to, e.g., one's school marks or chances of getting a job interview in the US. Therefore, we are interested in the influence of those profile information on the follower count.
We addressed this question by analyzing the profiles of about 6 Million Twitter users. All profiles are separated into three groups: Users that have a first name, English words, or neither of both in their name field. The assumption is that names and words influence the discoverability of a user and subsequently his/her follower count.
We propose a classifier that labels users who will increase their follower count within a month by applying different models based on the user's group. The classifiers are evaluated with the area under the receiver operator curve score and achieves a score above 0.800.

\end{abstract}

%
%

\ccsdesc[500]{Information systems~Probabilistic retrieval models}
\ccsdesc[500]{Information systems~Information extraction}
\ccsdesc[500]{Information systems~Clustering and classification}
\ccsdesc[300]{Computing methodologies~Machine learning}
\ccsdesc[300]{Human-centered computing~Social media}

\keywords{Prediction, Correlational Analysis, Classification, Social Network Analysis, Experimental Study, Onomastics}
\maketitle


\section{Introduction}
\label{sec:introduction}

Twitter is with 320 million monthly active users~\citep{twitter2015annual} one of the largest online social networks. It is important to follow the ``right'' users to maximize the user experience. But, what are the ``right'' users? Most scientists deem those users as ``right'' that produce a huge amount of interesting tweets---measured by their retweet, mention, and favorite rate---and those that have a huge follower count~\citep{gupta2013wtf,chorley2015human,razis2014influencetracker,aleahmad2015olfinder,hong2011predicting,hannon2010recommending,bigonha2012sentiment,bandari2012pulse,noro2013twitter,weng2010twitterrank}. This assumption is supported by a strong correlation between the number of tweets and the number of followers that are observed on Twitter by Beevolve\furl{http://beevolve.com/twitter-statistics} in October 2012 as shown in \figref{followers-vs-tweets}.

\fig{htpb}{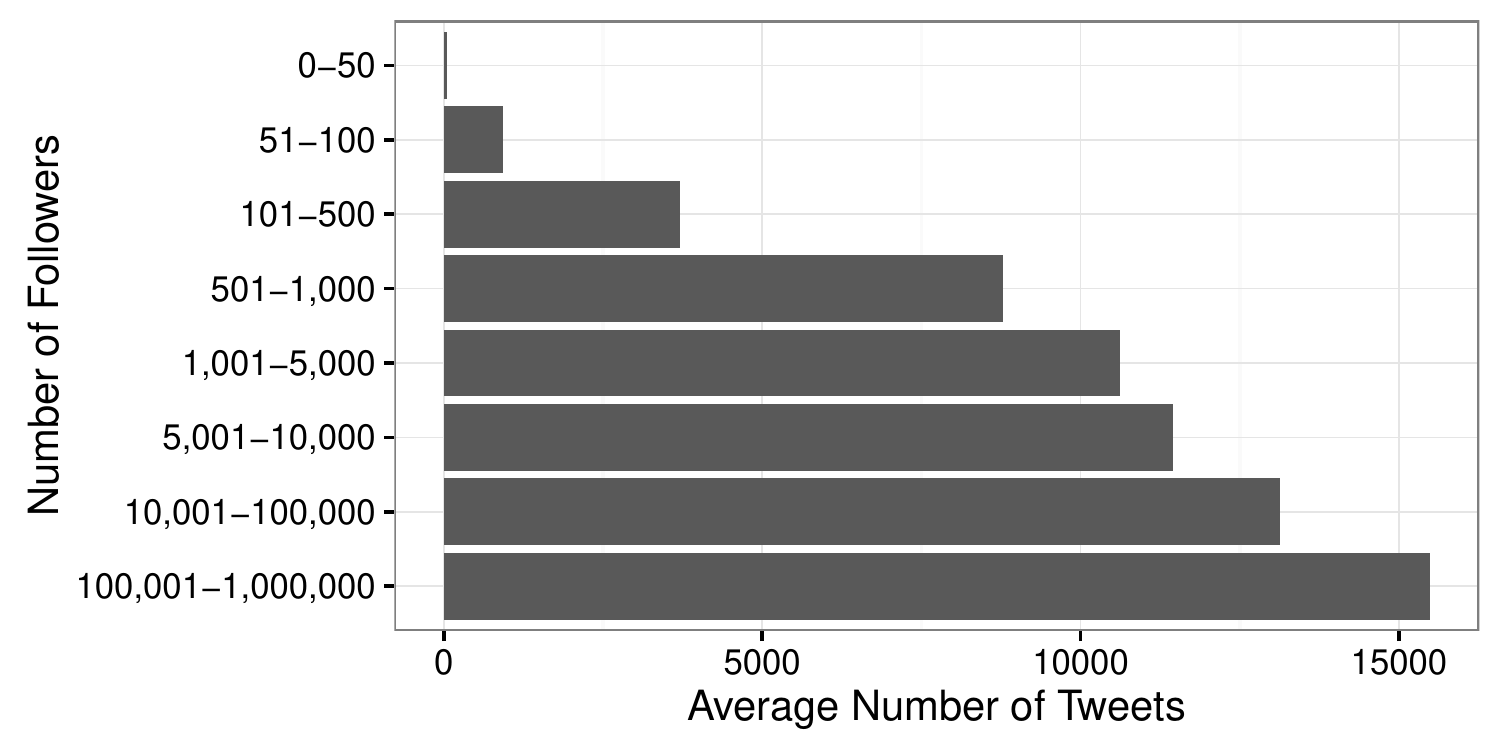}{Average number of tweets versus follower count (Source: Beevolve).}

With this assumption in mind, one could wonder whether these traits cover the manual discovery of new users to follow. Twitter offers only a text-based search to discover those users (recommended users aside). The search function focuses on the text information, given on a user's profile page. This means that a good selection of descriptive should increase visibility, because those terms are more likely to be entered in the search function.

The name is of interest, because studies in psychology and economics have identified correlations related to one's first name. Having a popular name helps to obtain good marks at school~\citep{kaiser2009vornamen}, while having a distinctively black name reduces your chances of getting a job interview~\citep{fryer2004causes,bertrand2003are}. Some names are strong signals for one's ideology and, consequently, could attract more likeminded people~\citep{oliver2015liberellas}. Therefore, our aim is to analyze the impact that the profile information as well as the content of name field have on the follower count. Understanding the influence of those information on the popularity can be used to increase the discoverability.

In this paper, we first identify the influence of the content of the real name field of a Twitter user has on her/his follower count. Then we identify those profile features that have the strongest relation to the increase of the follower count of that user. We will use those features in a novel user classifier that labels those users that are likely to gain more traction on Twitter---i.e., increase their follower count within a month. These users are of interest, because they could become the stars of tomorrow. The recommender utilizes only that information given in the user profile of the Twitter users. We do so, because of the following reasons:

\begin{enumerate}

	\item It reflects the way a user would scan the network for potential users to follow using the search feature rather than the timeline.

	\item It enables us to cover far more users than if we would include the status messages as well. The reason for this is that the Twitter API rate allows far more queries for different user profile information than it does for the timeline information of dedicated users. For instance, we could get all profile information from the dataset used for this paper within \daysForUsers~days---crawling the tweets would take about \monthsForTweets~months.

\end{enumerate}
Accordingly, we focus on those data that are accessible using the user profile information for answering the following research questions:

\begin{enumerate}

	\item What is the relation of the real name field and the popularity of a Twitter user?

	\item Which information from a user profile has the strongest relation to the popularity-gain of the users?

	\item Can we classify users that are about to increase their follower count?

\end{enumerate}
We will address each of those research questions in an experiment that will be described in this paper.

For this work, we refer to the content of the real name field on a profile page of a Twitter users as that users’ name. A Twitter user always has a second name that is used as human-readable unique identifier \seefig{twitter-profile}. This name is called username and is referred to in some literature as screen name. It is prefixed with an @-sign and not the content of our study.

\fig{htpb}{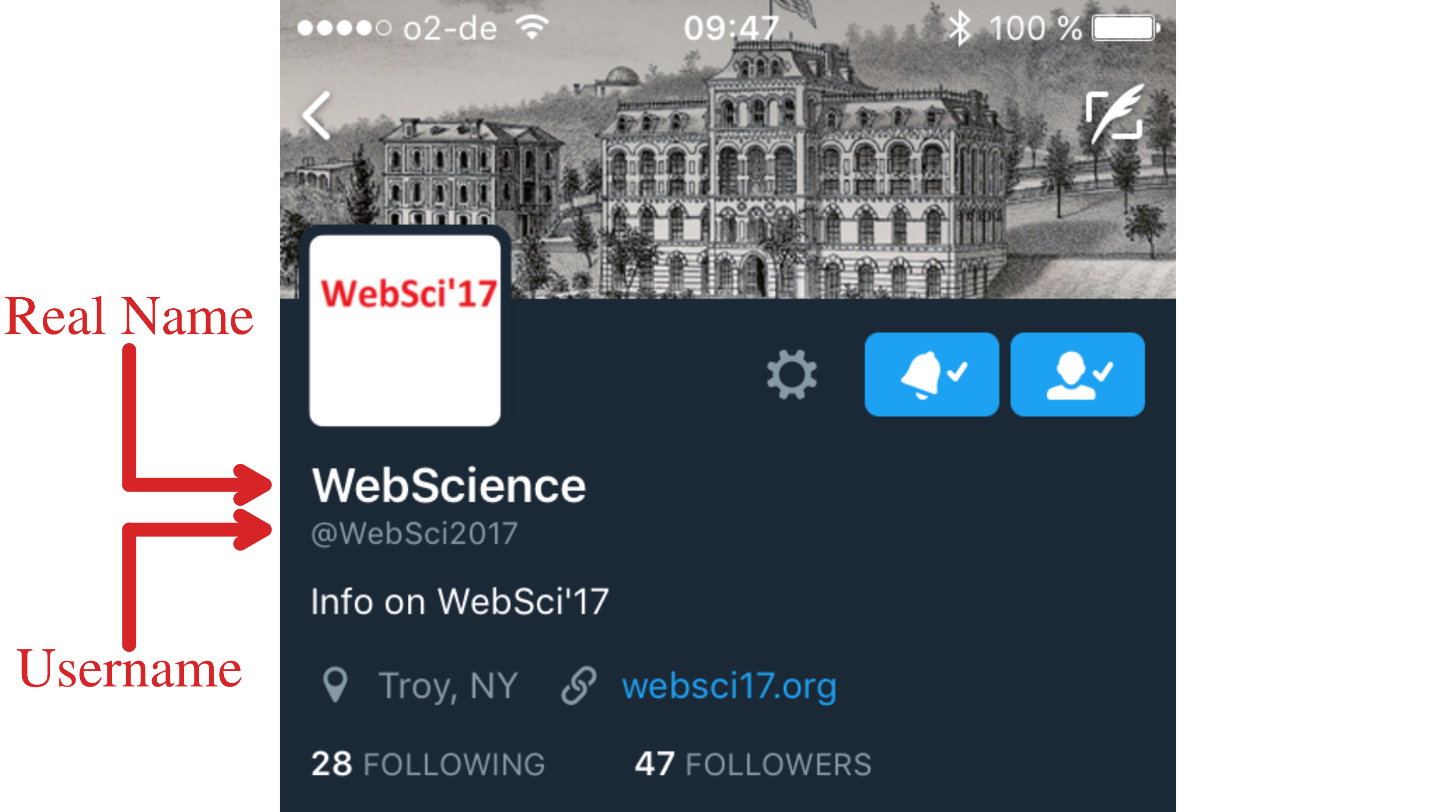}{Example of a Twitter profile, highlighting the real name (i.e., ``WebScience'') and the username (i.e., ``@WebSci2017'') of the user account of 9th International ACM Web Science Conference 2017.}

We propose an approach to label users that will or will increase their follower count within the next month. It uses a different classification model based on the content of the users' name. It differs between three types of user name contents: Contains a given name, contains English words, and contains neither of both. Each group's classifier can use a different model (e.g., Gradient Boosting Machine, Naive Bayes, etc.) that is trained with distinct parameters and features based on the corresponding group.
We can obtain classification results of above 0.800 across all groups as measured by the area under the receiver operator curve~(AUC) based on the classification probability.

The remainder of this paper is structured as follows: First, we will lay out the previous work on popularity-prediction as well as findings about given names in the following section. We then describe the used data of Twitter, English words, and names in the data section. Our first research question is addressed in \secref{experiment-1}, where we analyze the relation between the real name field and the follower count. The main contribution of this paper is presented in \secref{experiment-2} where we discuss our second research question, where we identify those profile features that have a statistical significant relation with an increase of the follower count. All findings are then applied in \secref{experiment-3}, where we present our classifier that assigns users to two categories: users that will increase their follower count and users that will not. Finally, we end our paper with the conclusions section.


\section{Related Work}
\label{sec:related-work}


Following \citet{riquelme2016measuring}, ``popularity'' describes the number of channels over which a user can reach other users~\citep{riquelme2016measuring}. Mostly, this refers to direct follow-up relationships, but extends to mentions of the author and replies to and retweets of the author's tweets. Popularity gets easily confused with the broader term ``influence''. Influential users are those users that can affect other users. Basically, it is beneficial for them to be popular (well connected) and have a decent amount of interesting output (i.e., tweets or retweets). Surprisingly, to the best of our knowledge, one channel is left out by the literature: lists. Twitter allows each user to create lists of users. A list can, for example, summarize all researches a given user knows on Twitter. They can be used to get a view on only the tweets of the users in the given list. It is worth mentioning that a user does not need to actively follow those users, i.e., it can contain tweets that are not part of the user's timeline. Therefore, we add this feature in our experiments.

In their survey on influence on Twitter, \citet{riquelme2016measuring} provide a good overview over influence metrics. The survey serves as toolkit to select the proper metrics for one's research. The authors describe eight different popularity measures. The simplest is to count only follow-up relations, but there are more complex ones, like the FollowerRank~\citep{nagmoti2010ranking,cappelletti2012iarank}, the Tweet-Follower-Followee ratio (TFF)~\citep{bigonha2012sentiment}, and the Popularity measure~\citep{aleahmad2015olfinder}. Those measures weight the follower count with the friend count. There are further metrics by \citet{srinivasan2014comparative} that consider the number of users that mention, reply to, or retweet the author's tweets.

However, FollowerRank, TFF, and the Popularity can be influenced by the given user by changing the number of users s/he follows. The measures by \citet{srinivasan2014comparative} are not feasible with only Twitter API access, because it would require crawling the mutual follow-relation of all users, which is very time-consuming. A detailed dump of Twitter data would be needed to get the required information. Further, \citet{riquelme2016measuring} write that---even with this level of access---computational issues can occur. Therefore, we will focus in this paper on the most simple measure, the follower count. It is not changeable by the given user and can be retrieved using the Twitter API within reasonable time-bounds.

The popularity of a Twitter user can be part of her/his influence. Consequently, the popularity is one factor to be used on influence tracking systems.
One such system is InfluenceTracker by \citet{razis2014influencetracker}. The authors describe the way their monitoring services ranks influential Twitter users by relying on the following characteristics: (1) TFF ratio: The authors argue that a user that follows more users than s/he has followers is a viewing user rather than a composing one. (2) Tweets Creation Rate~(TCR): The frequency of newly created tweets is added to identify more influential user if two users have the same TFF ratio. It is computed using the distance in hours between the 100 most recent tweets. (3) The authors present an h-index for retweets and Favorites. It is---as the TCR---calculated based on the 100 most recent tweets. This index should reflect the assessment of a user’s tweet by others. The authors use those three scores to compute an influence metric for every user. Amongst those, the TFF ratio takes the popularity of the user into account. However, as already mentioned at the popularity measures, this measure is under partial control of the user (s/he controls the number of friends).


\citet{suh2010want} conducted a large-scale analysis of Twitter data to identify those features that affect the retweet-ability of tweets the most. They considered the number of URLs, hashtags, and mentions of any given tweet as content feature and the number of followers and friends, the age in days of the account, the number of tweets, favorites, and retweets as contextual features. They discovered that the number of URLs and hashtags from the content features as well as the age and the number of follower and friends to be good predictors for the retweet-ability. Interestingly, the authors noted that the number of tweets was marginally a bad predictor. We will adopt some of their features for our analysis, but will also extend them by others that where left out by \citet{suh2010want}.

\citet{bandari2012pulse} went one step further and tried to predict the popularity of tweets. They focused on news stories and identified criteria using classification and regression methods that make a tweet more retweet-able. They selected as features the news category, the tone of language (i.e., subjective or objective), the mentioned users, as well as the source of the tweet. The authors were not able to predict the exact numbers of retweets using regression and obtained an accuracy of 85~\% using classification. They classified their data into three classes: low, medium, and high spread. This is a good decision, because this enables to reduce the effects of outliers. However, the authors give no information about the accuracy per class. It is likely that most accuracy is achieved in the biggest class for low spread. Nevertheless, we will adopt the classification approach used in this work and focus on classification, rather than regression for our predictor.

\citet{tsur2015dont} proposed an adapted Gradient Boosted Trees approach to predict the future popularity of hashtags. The authors extract several hashtag-related features from a huge dataset of tweets, like the length of the hashtag, the number of words in it, or the number of upper-cased letters. Their approach could obtain quite low error rates. However, as the previous approaches, we cannot transfer their finding, because it focusses solely on the tweets content.


\citet{hutto2013longitudinal} done a longitudinal study on follower count changes using a dataset of 507 users. Their study focuses on social behavior, message content, and network structure. They also included some profile information as the presence of a location or URL into their study. However, their sample is rather small and their focus lies more on the tweets than on the profile data.

\citet{martin2016hashtags} conducted a focused experimental study to isolate on the influence of the number of hashtags per tweet on the follower count. They used data from 502891 users and grouped them into a control and experimental group based on the presence of hashtags in the tweet. Their experiments show that users with hashtags increased their follower count by 2.88 and users without hashtags increased it by 0.88. However, like the study of \citet{hutto2013longitudinal}, their findings focus solely on the tweets.


In addition to current research, we want to add profile information as well as the name of the Twitter users to our experiments. To the best of our knowledge, no prior work has considered the influence of those features on a user's popularity. Those features are promising, given that for example \citet{gaudeul2015privacy} have discovered that disclosing the real name leads to smaller, but more productive networks.

\citet{kaiser2009vornamen} found out that teachers have negative prejudices regarding certain names like ``Kevin'' and ``Chantal'' in Germany.
Similar results were obtained in the United States by \citet{anderson-clark2008relationship} who found an effect for ethnic first names.
Such negative prejudices can have direct negative consequences on the student's evaluation and are most likely the effect of the halo effect \citep{allport1979nature,thorndike1920constant}. The halo effect is a cognitive bias that leads people to evaluate others based on their prejudices. An effect that is not limited to the classroom. Therefore, we will identify whether a un/popular name has on the popularity of a user.

Additionally, \citet{fryer2004causes} have tested the influence of distinctively black names on the economic outcome---that is one's chance to get a job. A distinctively black name was modeled using the share of blacks using a given name per birth certificates registered in California and Massachusetts. Prior findings suggested that distinctively black names are a disadvantage during job application and lead to higher rejection rates~\citep{bertrand2003are}. However, \citet{fryer2004causes} could not find negative effects of black names over white ones for later live outcomes after controlling the circumstances of birth. One of their explanations for their contradicting finding was that the disadvantage of black names holds only till the callback decision. A racist employer will not hire a black person, with or without a black name. However, the situation on Twitter is different, because a user's true identity is probably never revealed.

\citet{coffey2009do} have done a similar study where they conducted an empirical test with the hypothesis of whether masculine names help women to be successful in legal careers. The masculinity of a name was modeled by computing the share of men that use a given female name as well per South Carolina's voter registration dataset. They found a statistical significantly higher number of masculine names amongst South California Bar Association members; however, their finding is unable to pin down the source of this inequality.

Finally, \citet{oliver2015liberellas} found out that many parents try to signal their ideology by given certain names to their offspring using the same birth certificate data from California as \citet{fryer2004causes} did. It is consequently very likely that users with ideology carrying names (e.g., a liberal one) attracts other users with the same ideology, because of the homophily effect~\citep{mcpherson2001birds}, which describes a tendency of individuals to connect with similar ones.



\section{Data}
\label{sec:data}
We use three types of data to conduct our experiments: The Twitter dataset, a dataset to identify the presence of English words in the name field, and a dataset to identify given names in the same field. All three datasets are described in the following.

\subsection{Twitter Data}
We use the social graph dataset\fsource{http://an.kaist.ac.kr/traces/WWW2010.html} from \citet{kwak2010what} to get a list of user identifiers to crawl. The original dataset was collected between June 6th and June 31st of 2009 and contains all Twitter users at the time (41.7 million).
We crawled the users using their unique identifier with the Twitter API between September 12th and October 2nd of 2016. We could access only 30.5 of the 41.7 million users---a reduction of 26.6~\%. One reason for the decreased number of accessible users could be that the dataset was collected in 2009 and many users deactivated their account since then.

Amongst the collected users were 3760586 protected and 57087 verified profiles.
Protected users are users that have set their tweet privacy level to protected, meaning that only approved users can view their tweets. Users in this group do not contribute in the usual public way to the network and are likely to be followed by actual ``offline'' friends. A protected user must manually confirm every follower. Therefore, the follower counts of those users are different from the follower counts of ``regular users''.
Verified users are users that are confirmed by Twitter. Users in this group are already well known from outside of Twitter and, consequently, are being followed for their outside activity.
Therefore, we decided to remove both user groups from the sample.
We further removed any user account from our sample that has not sent any tweet within the past year. We consider those users as abundant---the owner has likely forgotten to delete his/her account.
These reductions lead to a remaining dataset of 3131421 users.

During our experiments, we assign every user to one of three groups based on the content of the real name field: ``Contains Name'', ``Contains Words'', and ``Custom Content'' (see Section~\ref{subsec:useddata}).
The descriptive statistics of the remaining user profiles in \tabref{descriptive-statistics-dataset} shows that the collected accounts are more than 7~years old. Surprisingly, the newest account was 334~days old, which should not be the case given that the dataset was originally collected in 2009. Obviously, Twitter reassigns old user IDs after some time. The collected users contributed with a median of 599~tweets to the social network. More than half of the users in the dataset have less than 112 followers, which indicates that we have a lot of non-prominent users in the dataset.
We further had 1899209 users that still use the default profile settings and 444045 use the default profile image (an egg on a colored background). 4127518 users provide a description, 4598818 a location, and 2241756 a URL.
As shown in \figref{follower-distribution}, the distribution of the follower counts in our data sample follows a power law distribution, which is in line with publications of Twitter researchers~\citep{gupta2013wtf}. The mean follow count is 792.251, while the median is at 50.


\begin{table*}[!t]
\centering
\caption{Descriptive statistics about the interval level measures of the dataset showing the mean, standard deviation~(SD), median, lower and upper bound of the 95~\% confidence interval~(CI), minimum~(Min), and maximum~(Max) value ($N = 6354052$).}
\label{tab:descriptive-statistics-dataset}
\begin{tabular}{lrrrr@{[}r@{, }rrr}
  \toprule
  Profile Feature           &     Mean &        SD & Median & \multicolumn{3}{r}{95~\% CI} & Min &      Max \\
  \midrule
  Followers Count           &  844.825 & 14263.897 &    112 &              &    5 &  2030] &   0 & 13359378 \\
  Friends Count             &  477.172 &  3166.262 &    194 &              &   12 &  1619] &   0 &  3106717 \\
  Tweet Count               & 4777.485 & 16056.563 &    599 &              &    5 & 22298] &   0 &  4345445 \\
  Favorited Count          &  794.268 &  4665.824 &     36 &              &    0 &  3262] &   0 &  1266503 \\
  Listed Count              &   16.698 &   955.009 &      2 &              &    0 &    63] &   0 &  2386908 \\
  Description URL Count     &    0.042 &     0.239 &      0 &              &    0 &     0] &   0 &        7 \\
  Description Hashtag Count &    0.125 &     0.737 &      0 &              &    0 &     0] &   0 &       30 \\
  UTC Offset                &   -2.411 &     4.502 &     -4 &              &   -7 &     8] & -11 &       13 \\
  Age in Days               & 2720.865 &    77.124 &   2713 &              & 2632 &  2888] & 334 &     3817 \\
  Inactivity in Days        &   67.801 &    96.327 &     15 &              &    0 &   295] &   0 &      365 \\
  \bottomrule
\end{tabular}
\end{table*}

\fig{htpb}{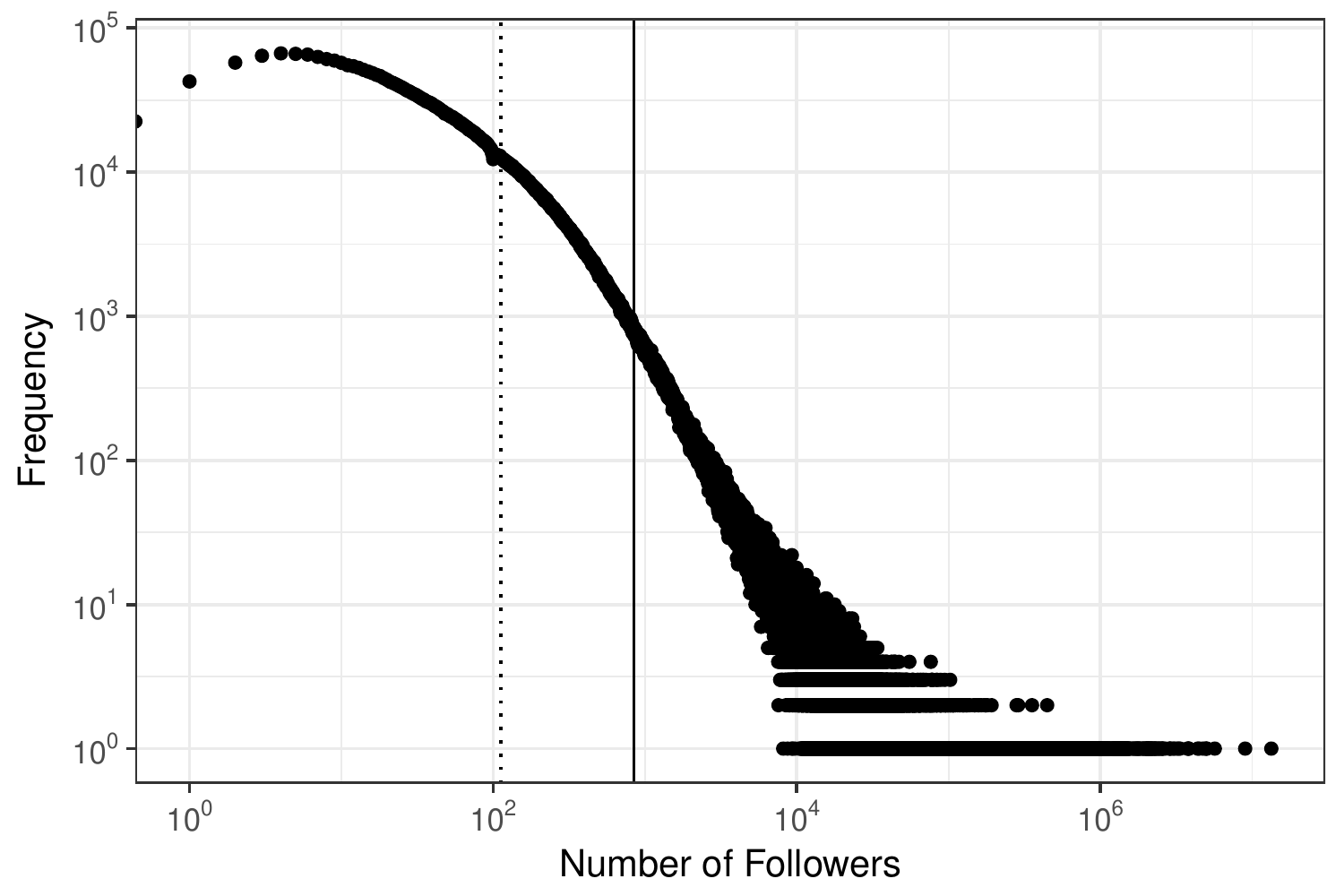}{Distribution of follower counts across all Twitter users (mean is solid line at 844.826; median is dotted line at 112).}

We need a second sample of the same users at a later time to have information about which users will increase their follower count. Therefore, we collected the same users again between November 10th and November 16th of 2016 to get information about their changed follower counts. We then compared the follower counts of both crawls and mark every user that could improve his/her follower count during that time with a flag. This resulted in 2789168 (43.9~\%) flagged users that improved their follower count. \tabref{descriptive-statistics-exp2} shows some descriptive statistics about the follower count of both crawls as well as the change of it.


\begin{table*}[tpb]
\centering
\caption{Descriptive statistics about the follower counts accross the two samples ($N = 6354052$).}
\label{tab:descriptive-statistics-exp2}
\begin{tabular}{lrrrr@{[}r@{, }rrr}
	\toprule
	Follower Count  &    Mean &        SD & Median & \multicolumn{3}{r}{95~\% CI} &         Min &          Max \\
	\midrule
  First Dataset   & 844.825 & 14263.897 & 112.000 &        &  5.000 & 2030.000] &       0.000 & 13359378.000 \\
  Second Dataset  & 856.920 & 14480.515 & 113.000 &        &  6.000 & 2058.000] &       0.000 & 13395833.000 \\
	\midrule
  Absolute Change &  12.095 &   500.767 &   0.000 &        & -5.000 &   23.000] & -126099.000 &   437959.000 \\
  Relative Change &   0.989 &     0.196 &   1.000 &        &  0.917 &    1.032] &       0.000 &      315.000 \\
	\bottomrule
\end{tabular}
\end{table*}

\subsection{English Wordlist}
We used the English wordlist\fsource{http://sil.org/linguistics/wordlists/english} from SIL International to identify the presence of English words in the name field of a Twitter users. It is a collection of 109582 English words and was originally obtained from the Interocitor bulletin board in Dallas (214-258-1832).

\subsection{Name Data}
The text in a Twitter user's name field where converted to ASCII, lower-cased, separated by non-letters, and then matched against the names that are provided in Behind the Name.\footnote{Campbell, Michael D. ``Behind the Name: the Etymology and History of First Names.'' http://behindthename.com} It is a website that tries to colect all names in order to make them accessible for the public.
We collected \DataNamesNum distinct names from which \DataNamesNumMale are male, \DataNamesNumFemale are female, and \DataNamesNumUnisex are used by both genders.
Additionally, we collected data about name impressions to cover the subjective impressions of names. These data are important to check for effects like negative prejudices for certain names as in the findings of \citet{kaiser2009vornamen}. They were created using the feedback of the Behind the Name users and consists of the following data (counts are given in the parentheses): good vs. bad (7666/1390), masculine vs. feminine (6325/6471), classic vs. modern (8127/1771), mature vs. youthful (5283/3624), formal vs. informal (6473/3080), upper-class vs. common (6640/2125), urban vs. natural (2076/6054), wholesome vs. devious (7651/1387), strong vs. delicate (8143/1523), refined vs. rough (7070/2316), strange vs. boring (11286/164), simple vs. complex (4037/5214), serious vs. comedic (7328/1679), and nerdy vs. unintellectual (5664/1577). A name that is present in one of those list is associated by the Behind the Name community with the respective trait.


\section{Experiment 1 -- Effects of Names on Popularity}
\label{sec:experiment-1}
Our first experiment focuses on the question of whether the presence of given names or English words in the real name field of a Twitter users profile improves his or her popularity. Having well known words or names in there should increase the discoverability of a user and, therefore, his/her chance of more followers.


\begin{table*}[tpb]
\centering
\caption{Descriptive statistics about the follower counts accross the groups.}
\label{tab:descriptive-statistics-anova}
\begin{tabular}{lrrrrr@{[}r@{, }rrr}
  \toprule
  User Group      &       N &     Mean &        SD & Median & \multicolumn{3}{r}{95~\% CI} & Min &     Max \\
  \midrule
  Custom Content  & 1495530 & 1040.749 & 19476.297 &    130 &                  & 6 & 2413] &   0 & 13359378 \\
  Contains Words  & 471312  & 2320.123 & 24841.446 &    289 &                  &10 & 7420] &   0 &  8988128 \\
  Contains a Name & 4387210 &  619.549 &  9938.525 &     97 &                  & 5 & 1503] &   0 &  5670985 \\
  \midrule
  Overall         & 6354052 &  844.826 & 14263.900 &    112 &                  & 5 & 2030] &   0 & 13359378 \\
   \bottomrule
\end{tabular}
\end{table*}

\subsection{Used Data}
\label{subsec:useddata}
We want to measure the effect of the name field on the follower count of a Twitter user. Therefore, we will parse the name fields and assign every Twitter user to one of three groups:
\begin{enumerate}

	\item Contains Name: This group contains users that have a given name in their name field that we could match to the Behind the Name data.

	\item Contains Words: This group contains all users that are not in the first group, but have at least one English word from the SIL list in their name field.

	\item Custom Content: This group contains all users that are neither in the first nor in the second group.

\end{enumerate}

\tabref{descriptive-statistics-anova} shows some descriptive statistics about the follower count for all three groups. Users with given names have the fewest followers with a median 97 and a mean of 619.549.
Users that have English words in their real name field are---with a median of 289 and a mean of 2320.123---the most followed group by far.
Users from the ``Custom Content'' group are---with a median of 130 and a mean of 1040.749---between the other two groups

\subsection{Design and Procedure}
We will conduct a one-way analysis of variance~(ANOVA) on the follower count of the users from the ``Contains Name'', ``Contains Word'', and ``Custom Content'' groups using SPSS version 24. We want to find out whether there is a statistical difference between these groups regarding their follower count.

Further, we want to identify whether there is a statistically significant difference between the follower counts of each group. Therefore, we will add a Tukey post-hoc test to find the differences between the groups if the ANOVA finds a difference in the groups.

\subsection{Results}
The one-way ANOVA results presented in \tabref{exp1:anova-result} show that there is a statistically significant difference between our experimental groups ($F(2,6354049) = 3212.414, p = 0.000$).

A Tukey HSD post hoc test reveals that the followers count is statistically significantly higher for users with English words ($2320.123 \pm 24841.446, p = 0.000$) and lower for those with given names in their name field ($619.549 \pm 9938.525, p = 0.000$) compared to the users with custom content in their name field ($1040.749 \pm 19476.297$). There is a statistically significant difference between those users with English words and given names in their name field as well ($p = 0.000$).

\begin{table}[hbp]
\begin{threeparttable}
	\centering
	\caption{Results from the analysis of variance~(ANOVA) for the followers count showing the sum of squares~(SS), degrees of freedom~(df), mean square~(MS), and F-value.}
	\label{tab:exp1:anova-result}
	\begin{tabular}{lrrrr}
		\toprule
		Source  &       SS &      df &       MS &        F \\
		\midrule
		Between & 1.306e12 &       2 & 6.529e11 & 3212.414 \\
		Within  & 1.291e15 & 6354049 & 2.033e08 &          \\
		Total   & 1.293e14 & 6354051 &          &          \\
		\bottomrule
	\end{tabular}
    \begin{tablenotes}
      \small
    \item Significant at $p<0.001$.
    \end{tablenotes}
\end{threeparttable}
\end{table}

\subsection{Discussion}
Twitter uses the name field as part of its user search feature.\furl{https://dev.twitter.com/rest/reference/get/users/search} Our hypothesis is that having known terms in there should consequently increase the chance to be found by it. Consequently, having English words and given names makes it easier to be found using text search. However, this should not directly translate in huge follower counts as \citet{gaudeul2015privacy} pointed out: Disclosing your given names results in smaller, but more profitable networks.

This hypothesis is confirmed by the one-way ANOVA with Tukey HSD post-hoc test. Users that have English words in their name field have statistically significant higher follow counts. On the other hand, users that have a given name in their name field have less followers, which is in line with the finding of \citet{gaudeul2015privacy}. A difference that will implicitly be used by our later experiments where we want to predict an increase of the follower count based on the content of the real name field.

The high follower counts of users with ``Custom Content'' can be explained by other languages---our experiments only considers English words. Therefore, ``Custom Content'' refers not only to misspelled or Emoji names, but could include valid words from other languages as well. Those in turn, are easy to discover by the search function.

While we observed statistically significant differences in the follower counts amongst the groups, we do not know where they come from. However, the presence of this difference could be used to predict popular users. The feature is easy to extract from the user profile data and can serve as a feature in a popularity predictor.


\section{Experiment 2 -- Identifying Predictive Features}
\label{sec:experiment-2}
Our second experiment tries to identify user profile features that have the strongest statistical relation with the change of the follower count. In other words, what features are the best proxy to predict an increase of the follower count.

\subsection{Used Data}
During this experiment, we used the second crawl of the users to access the information about their change of the follower count. Further, we add the subjective name features from Behind the Name to the users from the ``Contains Name'' group to identify the influence of the perception of a given name to the follower count. Additionally, we add the share of English words in the real name field to all users except of those from the ``Custom Content'' group.

We mark with a flag, called \emph{follower count increased flag} in the sequel, all users with an increased follower count in the second crawl.
Then the dataset is randomly split into halves while preserving the relative ratio of the follower count increased flag. We will use the first partition to identify features with a statistical significant effect on the flag during this experiment---the second partition will be used later in the following experiment.

\subsection{Design and Procedure}
We will conduct a binary logistic regression on the follower count increased flag for users in the ``Contains Name'', ``Contains Word'', and ``Custom Content'' groups separately using SPSS version 24.

We want to identify which profile features have a statistical significant relation with the follower count increased flag. The results from our first experiment showed that users from all three groups show a different pattern in their follower count. Therefore, we want to identify the features for every group separately. We will use this detailed information in our third and last experiment where we present our popularity predictor.

\subsection{Results}
A binary logistic regression for every user group was performed to identify the effects of all user profile features as well as the Behind the Names data on the follower count changed flag.

The variation in the follower count that can be explained by the models are presented in \tabref{exp2:model-summary}. Therein, our models can explain about 21.0~\% (Nagelkerke R$^2$) of the variation of users with custom content, 23.8~\% for users with English word in their name, and 20.5~\% for users with names in their real name field.

\begin{table}[htpb]
    \centering
    \caption{Amount of variation in the dependent variables that can be explained by the models of every user group.}
    \label{tab:exp2:model-summary}
    \begin{tabular}{lr}
        \toprule
        User Group     & Nagelkerke R$^2$ \\
        \midrule
        Custom Content &           0.210 \\
        Contains Words &           0.238 \\
        Contains Names &           0.205 \\
        \bottomrule
    \end{tabular}
\end{table}

The effect sizes are presented in \tabref{exp2:regression-results}. We limited the presented variables to those that are statistically significant for space reasons.

\begin{table}[htpb]
\centering
\begin{threeparttable}
    \caption{The contribution of each independent variable to the models and its statistical significance as determined by a Wald test. The results are separated by user group~(G) and show the significance as well as the $\beta$ value, its standard error~(SE), and the effect size ($e^\beta$).}
    \label{tab:exp2:regression-results}
    \begin{tabular}{c@{\enskip}lr@{}lrr}
        \toprule
        G & Variable                & $\beta$&         &    SE & $e^\beta$ \\
        \midrule
        \multirow{13}{*}{\rotatebox{90}{Custom Content}}

    & Age in Days                &  0,000&$^{***}$ & 0,000 & 1,000 \\
    & Inactivity in Days         & -0,009&$^{***}$ & 0,000 & 0,991 \\
    & Tweet Count                &  0,000&$^{**}$  & 0,002 & 1,000 \\
    & Favorited Count           &  0,000&$^{***}$ & 0,000 & 1,000 \\
    & Friends Count              &  0,000&$^{***}$ & 0,000 & 1,000 \\
    & Listed Count               &  0,002&$^{***}$ & 0,000 & 1,002 \\
    & Description URL Count      &  0,036&$^{***}$ & 0,000 & 1,037 \\
    & Description Hashtag Count  &  0,013&$^{***}$ & 0,000 & 1,013 \\
    & Has Default Profile        &  0,102&$^{***}$ & 0,000 & 1,107 \\
    & Has Default Profile Image  & -0,872&$^{***}$ & 0,000 & 0,418 \\
    & Has Description            &  0,214&$^{***}$ & 0,000 & 1,238 \\
    & Has Location               &  0,099&$^{***}$ & 0,000 & 1,104 \\
    & Has URL                    &  0,221&$^{***}$ & 0,000 & 1,247 \\
        \midrule
        \multirow{14}{*}{\rotatebox{90}{Contains Words}}
    & Age in Days                &  0,000&$^{***}$ & 0,000 & 1,000 \\
    & Inactivity in Days         & -0,010&$^{***}$ & 0,000 & 0,990 \\
    & Tweet Count                &  0,000&$^{***}$ & 0,000 & 1,000 \\
    & Favorited Count           &  0,000&$^{***}$ & 0,000 & 1,000 \\
    & Friends Count              &  0,000&$^{***}$ & 0,000 & 1,000 \\
    & Listed Count               &  0,003&$^{***}$ & 0,000 & 1,003 \\
    & Description URL Count      & -0,032&$^{**}$  & 0,002 & 0,969 \\
    & Description Hashtag Count  & -0,010&$^{**}$  & 0,002 & 0,990 \\
    & Has Default Profile        &  0,098&$^{***}$ & 0,000 & 1,103 \\
    & Has Default Profile Image  & -0,889&$^{***}$ & 0,000 & 0,411 \\
    & Has Description            &  0,223&$^{***}$ & 0,000 & 1,250 \\
    & Has Location               &  0,088&$^{***}$ & 0,000 & 1,092 \\
    & Has URL                    &  0,445&$^{***}$ & 0,000 & 1,561 \\
    & Name Countains Words       & -0,187&$^{***}$ & 0,000 & 0,829 \\
        \midrule
        \multirow{24}{*}{\rotatebox{90}{Contains Names}}
    & Age in Days                &  0,001&$^{***}$ & 0,000 & 1,001 \\
    & Inactivity in Days         & -0,009&$^{***}$ & 0,000 & 0,991 \\
    & Tweet Count                &  0,000&$^{***}$ & 0,000 & 1,000 \\
    & Favorited Count           &  0,000&$^{***}$ & 0,000 & 1,000 \\
    & Friends Count              &  0,000&$^{***}$ & 0,000 & 1,000 \\
    & Listed Count               &  0,001&$^{***}$ & 0,000 & 1,001 \\
    & Description URL Count      &  0,055&$^{***}$ & 0,000 & 1,056 \\
    & Description Hashtag Count  &  0,029&$^{***}$ & 0,000 & 1,029 \\
    & Has Default Profile        &  0,120&$^{***}$ & 0,000 & 1,128 \\
    & Has Default Profile Image  & -0,898&$^{***}$ & 0,000 & 0,407 \\
    & Has Description            &  0,248&$^{***}$ & 0,000 & 1,281 \\
    & Has Location               &  0,055&$^{***}$ & 0,000 & 1,056 \\
    & Has URL                    &  0,161&$^{***}$ & 0,000 & 1,175 \\
    & Name is Male               & -0,011&$^{*}$   & 0,023 & 0,989 \\
    & Name is Female             & -0,058&$^{***}$ & 0,000 & 0,944 \\
    & Impression: bad            & -0,027&$^{**}$  & 0,001 & 0,973 \\
    & Impression: classic        &  0,010&$^{**}$  & 0,001 & 1,010 \\
    & Impression: comedic        & -0,011&$^{**}$  & 0,007 & 0,989 \\
    & Impression: devious        &  0,022&$^{**}$  & 0,008 & 1,022 \\
    & Impression: feminine       & -0,058&$^{***}$ & 0,000 & 0,944 \\
        \multicolumn{5}{c}{-- Continued on next page --} \\
    \end{tabular}
\end{threeparttable}
\end{table}

\begin{table}[htpb]
\centering
\begin{threeparttable}
    \begin{tabular}{c@{\enskip}lr@{}lrr}
        \multicolumn{5}{c}{{\tabref{exp2:regression-results} -- Concluded from previous page}} \\
        \\
        \toprule
        G & Variable                 & $\beta$&        &    SE & $e^\beta$ \\
        \midrule
        \multirow{13}{*}{\rotatebox{90}{Contains Names}}
    & Impression: formal         &  0,011&$^{**}$  & 0,001 & 1,011 \\
    & Impression: good           & -0,026&$^{***}$ & 0,000 & 0,975 \\
    & Impression: masculine      & -0,033&$^{***}$ & 0,000 & 0,968 \\
    & Impression: mature         & -0,010&$^{**}$  & 0,001 & 0,990 \\
    & Impression: modern         & -0,032&$^{***}$ & 0,000 & 0,969 \\
    & Impression: nerdy          &  0,011&$^{***}$ & 0,000 & 1,011 \\
    & Impression: rough          & -0,019&$^{***}$ & 0,000 & 0,981 \\
    & Impression: serious        &  0,015&$^{***}$ & 0,000 & 1,015 \\
    & Impression: strange        &  0,012&$^{***}$ & 0,000 & 1,012 \\
    & Impression: strong         &  0,007&$^{*}$   & 0,028 & 1,007 \\
    & Impression: unintellectual &  0,013&$^{*}$   & 0,016 & 1,014 \\
    & Impression: youthful       &  0,009&$^{*}$   & 0,015 & 1,009 \\
    & Name Contains Words       & -0,068&$^{***}$ & 0,000 & 0,934 \\
        \bottomrule
    \end{tabular}
  \begin{tablenotes}
    \item $^{***}$            Significant at $p<0.001$
    \item $^{**\hphantom{*}}$ Significant at $p<0.010$
    \item $^{*\hphantom{**}}$ Significant at $p<0.050$
  \end{tablenotes}
\end{threeparttable}
\end{table}

The logistic regression model for users with custom content in their name contains all variables. The biggest effects come from two distinct areas: The degree of customization of the user profile and the presence of additional meta information.
On the one hand, the likelihood of additional followers increases if a user uses the default profile settings ($+10.7~\%$), provides a description ($+23.8~\%$), location ($+10.4~\%$), or URL ($+24.7~\%$).
On the other hand, the likelihood of additional followers decreases dramatically if the user did not provide a custom profile image ($-58.2~\%$).

The logistic regression model for users with English words in their name contains all possible variables as well.
In this group, the likelihood of additional followers increases if a user uses the default profile settings ($+10.3~\%$), provides a description ($+25.0~\%$), location ($+9.2~\%$), or URL ($+56.1~\%$).
The likelihood of additional followers decreases dramatically as well if the user did not provide a custom profile image ($-58.9~\%$).
Further, the fraction of English words in the name field decreases the likelihood that the follower count increases by $17.1~\%$.

The logistic regression model for users with English words in their name contains all possible profile information variables. Amongst the statistical not significant ones are the impressions boring ($p = 0.348$), common ($p = 0.982$), complex ($p = 0.762$), delicate ($p = 0.151$), informal ($p = 0.176$), natural ($p = 0.438$), refined ($p = 0.088$), simple ($p = 0.303$), upper-class ($p = 0.264$), urban ($p = 0.207$), and wholesome ($p = 0.378$).
In this group, the likelihood of additional followers increases if a user uses the default profile settings ($+12.8~\%$), provides a description ($+28.1~\%$), location ($+5.6~\%$), or URL ($+17.5~\%$).
The absence of a custom profile image ($-58.3~\%$) as well as the fraction of English words ($-6.6~\%$) in the real name field decreased the likelihood of additional followers.
The biggest changes amongst the name-related features are obtained for names that are perceived as female ($-5.6~\%$) or give a feminine impression ($-5.6~\%$).

\subsection{Discussion}
In \secref{related-work}, we pointed out that there is one user-relation missing in the current literature: the listed count. So, how does the list count performs in our analysis. Not so well. It is a statistical significant variable and its effect size is higher than those of the follower, friend, and tweet count in all three groups. However, the effect size is still quite low.

We can see some analogies, while comparing our results with those from tweet-based popularity predictions (those that extend their variables to profile information). \citet{suh2010want} found a statistical significant effect for follower and friends count, the age of the account in days, and the presence of a URL, which is in line with our results. However, in our results, only the presence of a URL has a relevant effect size (different from 1.000).
\citet{hutto2013longitudinal} conducted another study involving profile information. They identified the description length as well as the presence of a URL and a location. We have not included the length of the description into our model, but its presence. All three variables are indeed statistical significant and have a relevant effect size in our models.

Our analysis also shows that the subjective impression of a name matters. In most cases the effects are rather small, except for names that are perceived as ``female'' or ``feminine''. Those perceptions reduce the likelihood of additional followers. This indicates a slight discrimination of women on Twitter, probably comparable to those found by \citet{coffey2009do} in legal careers. Our data allow no comparison with the findings about distinctively black names; however, we have some information about names that are perceived as bad or rough (with a decreasing effect) and names that are perceived as strange or unintellectual (with an increasing effect). Especially the later ones indicate that negative prejudices about names do not necessarily translate into direct negative consequences.

Our analysis show that we can model the flag about the increase of the follower count for each user group using a different set of variables. All user groups share a strong core of five variables, consisting of the flags for default profile and profile image as well as the presence of a description, location, and URL. The fraction of English word in the name field is also a good additional predictor for users of the ``Contains Words'' group. Users from the ``Contains Names'' group extend the list of predictors by some name related variables.

\section{Experiment 3 -- Classifying Users}
\label{sec:experiment-3}
Our final experiment focuses on the question of whether we can correctly predict if the follower count of a Twitter user will increase based on her/his profile information within a time-span of one month.
We will predict only if the follower count of a Twitter user will increase. It will not reflect the magnitude of the change. The work of \citet{bandari2012pulse} already shows that a regression model is not very promising. Further, there is no scientific way to set clear ranges.

\subsection{Used Data}
We use the same data that was used during experiment~2. The first partition of the data will be used to find good parameters for every model during a grid search approach and to train the best model of each group using the found parameters.
The second partition will we used in the second part of this experiment to evaluate the performance of the trained models.

\subsection{Design and Procedure}
We conducted our experiments with GNU R 3.3.3\furl{https://r-project.org} using the caret package in version 6.0-73.\furl{https://github.com/topepo/caret} We test a broad selection of classification models that are available in caret: C5.0, Gradient Boosting Machine, k-Nearest Neighbors, Naive Bayes, Neural Network, and Random Forest.
We start by conducting a grid search approach on all models using the first partition of the user data. The best configuration of each model is used to compare all models against each other to select the best model for every group. We then train those selected models using the first partition of the data and use them to classify the data in the second partition of the dataset.
Our classifier then first identifies the group a given user belongs to by analyzing the content of the real name field. The model that was trained for the assigned group is then used to classify the user.

The classification will be evaluated using the area under the receiver operator curve (AUC) with a 10-fold cross-validation with five repeats. We will compare this approach with a general solution that applies the same model on all data as well as a random classifier that serves as a lower baseline.

\subsection{Results}
\tabref{exp3:auc-all} shows the results from the grid search. Shown are only the parameter and AUC scores of the best found configuration.


\begin{table*}[htpb]
\centering
\caption{Detailed results for all prediction models on every group showing the best found parameter as well as the accuracy under the receiver operator curve (AUC) value.}
\label{tab:exp3:auc-all}
\begin{tabular}{cllr}
  \toprule
  Group & Model & Parameter & AUC \\
  \midrule
  \multirow{6}{*}{\rotatebox{90}{Custom Content}}
  & C5.0 & boosting iterations: 25;  model: tree; winnowing: FALSE & 0.809 \\
  & Gradient Boosting Machine & boosting iterations: 1600; maximux tree depth: 15; shrinkage: 0.025 & 0.815 \\
  & k-Nearest Neighbors & k: 135 & 0.764 \\
  & Naive Bayes & Laplace correction: FALSE; kernel: TRUE; bandwidth adjustment: TRUE & 0.761 \\
  & Neural Network & hidden units: 11; decay weigth: 1.1 & 0.762 \\
  & Random Forest & randomly selected predictors: 3 & 0.808 \\
  \midrule
  \multirow{6}{*}{\rotatebox{90}{Contains Words}}
  & C5.0 & boosting iterations: 25;  model: rules; winnowing: FALSE & 0.833 \\
  & Gradient Boosting Machine & boosting iterations: 1100; maximux tree depth: 15; shrinkage: 0.025 & 0.838 \\
  & k-Nearest Neighbors & k: 100 & 0.762 \\
  & Naive Bayes & Laplace correction: FALSE; kernel: TRUE; bandwidth adjustment: TRUE & 0.782 \\
  & Neural Network & hidden units: 3; decay weigth: 1.2 & 0.775 \\
  & Random Forest & randomly selected predictors: 3 & 0.832 \\
  \midrule
  \multirow{6}{*}{\rotatebox{90}{Contains Names}}
  & C5.0 & boosting iterations: 25;  model: tree; winnowing: FALSE & 0.806 \\
  & Gradient Boosting Machine & boosting iterations: 1600; maximux tree depth: 15; shrinkage: 0.025 & 0.813 \\
  & k-Nearest Neighbors & k: 135 & 0.777 \\
  & Naive Bayes & Laplace correction: FALSE; kernel: TRUE; bandwidth adjustment: TRUE & 0.755 \\
  & Neural Network & hidden units: 13; decay weigth: 1.1 & 0.773 \\
  & Random Forest & randomly selected predictors: 7 & 0.805 \\
  \bottomrule
\end{tabular}
\end{table*}

The parameters of some models seem to be equal across all groups as the Naive Base without Laplace correction, kernel, and bandwidth adjustment. Others do have quit similar configurations across all groups as the C5.0 classifier which has 25 boosting iterations, no winnowing, but with differences in the used model or the Gradient Boosting Machine with a tree depth of 15, a shrinkage of 0.025, but a different number of boosting iterations.
The obtained AUC scores are in a range between 0.761 by the Naive Bayes on the ``Custom Content'' group and 0.838 by the Gradient Boosting Machine on the ``Contains Words'' group. k-Nearest Neighbors, Naive Bayes, and Neural Network perform bad across all groups with AUC scores bellow 0.800, while the C5.0, Gradient Boosting Machine, and Random Forest perform comparable well in all groups with Gradient Boosting Machine being consistently the best model.

Therefore, a Gradient Boosting Machine will be used during our evaluation for all groups, but it will differ in the used parameters and used portion of the train data.


\begin{table}[btb]
\centering
\caption{Overall results for all prediction models using the accuracy under the receiver operator curve (AUC).}
\label{tab:exp3:auc}
\begin{tabular}{llr}
  \toprule
  Group & Model & AUC \\
  \midrule
  Custom Content & Gradient Boosting Machine & 0.815 \\
  Contains Names & Gradient Boosting Machine & 0.812 \\
  Contains Words & Gradient Boosting Machine & 0.838 \\
  \midrule
  Overall &  & 0.816 \\
  \bottomrule
\end{tabular}
\end{table}

\tabref{exp3:auc} shows the AUC results of the final classification task. A Gradient Boosting Machine was used for every group, using a different parameter set as well as a different feature set, based on the assigned group.
The best score was obtained with 0.838 for users in the ``Contains Words'' group, followed be the ``Custom Content'' group with 0.815, and the ``Contains Name'' group with 0.812. The group-wise scores are in line with the scores from the training stage in \tabref{exp3:auc-all} and essentially the same for all groups except of a tiny difference of 0.001 in the ``Contains Names'' group.

Given the low variance between those results, one might raise the question of whether the distinction of the three groups brings any benefits. However, using the most general model from the ``Custom Content'' group for all data results in lower AUC scores for the remaining groups (i.e., 0.813 for ``Contains Words'' and 0.812 for ``Contains Names'').

\subsection{Discussion}

Our proposed model performs well on all three experimental groups. We were able to classify especially users from the ``Contains Words'' group better than those from the other groups.

We see a clearly better outcome while comparing our results with the random baseline, which would obtain an AUC score of 0.500 across all groups. Therefore, we can say that our classifier is far better than if we would guess which user will increase his/her follower count by random. However, we should keep in mind that the random classifier is a poor classifier and can serve only as a lower bound.

We can further compare our approach to a simple model that classifies all users with the same model---and by doing so, ignoring all additional name field-related features for the model. This comparison leads to some interesting results. The classification of users in the ``Contains Words'' group obtained far better scores with our approach, which lead to a reduction of the classification error rate of 13.4~\%. This group benefits the most from the name field-based distinction of the groups. However, we can see no relevant improvement for users in the ``Contains Names'' group. This is surprising, given that this model had a bunch of additional statistical significant variables with a decent effect size to draw from~\seetab{exp2:regression-results}. We should spend more research in finding good ways to use those effects for our classification.


\section{Conclusions}
\label{sec:conclusion}

A key feature of our proposed approach is the incorporation of the content of the name field, which was to the best of our knowledge never used before. We used the presence of English words and given names to form three groups (i.e., ``Custom Content'', ``Contains Words'', and ``Contains Names''). We combined the found names with meta-information from Behind the Name to add more details about those names. Our experiments show that adding those features that stem from the content of the name field leads to better outcomes for the prediction of an increase in follower counts.

Our experimental study identified multiple predictors from the profile information of a Twitter user that can be used to predict an increase of the follower count. Focusing on variables that are given in the profile information enables far bigger studies than those that use the content of the tweets as it is the case in the studies of \citet{hutto2013longitudinal} and \citet{martin2016hashtags}---because of the rate limitations of the Twitter API.

One limitation of our approach is that it focusses solely on words of the English language. Users with words from other languages are likely to be assigned to the ``Custom Content'' group, which is especially interesting given the spread of languages like Spanish, which is also a very common language on Twitter. One could try to evaluate the ``lang''-Field of the Twitter profile, which conveys the self-declared user interface language. However, most users in our data have set their interface to English with only a marginally difference for users in the ``Custom Content'' group (i.e., 81.12~\% in the ``Custom Content'', 90.39~\% in the ``Contains Words'', and 89.98~\% in the ``Contains Names'' group). Therefore, it would be more promising to detect the language of a user using machine learning in order to make further distinctions on the language.

Another limitation of our approach is that it predicts only if a user increases her/his follower count. Further research is needed to find a way to map the change into classes.

Finally, we would like to point out that our model is based on statistical differences and we do not yet have a clear indication where they come from. Our hypothesis is that more familiar content in the user profiles lead to better discoverability. For instance, the effect size of the default profile flag indicates that users with default profiles (without a description, location, URL, or profile image) are seen a less invested in Twitter and, therefore, less interesting to follow. Therefore, it would be of interest to conduct further social research to identify those causes.

\bibliographystyle{ACM-Reference-Format}
\bibliography{popularity-references}


\begin{thebibliography}{00}


\ifx \showCODEN    \undefined \def \showCODEN     #1{\unskip}     \fi
\ifx \showDOI      \undefined \def \showDOI       #1{#1}\fi
\ifx \showISBNx    \undefined \def \showISBNx     #1{\unskip}     \fi
\ifx \showISBNxiii \undefined \def \showISBNxiii  #1{\unskip}     \fi
\ifx \showISSN     \undefined \def \showISSN      #1{\unskip}     \fi
\ifx \showLCCN     \undefined \def \showLCCN      #1{\unskip}     \fi
\ifx \shownote     \undefined \def \shownote      #1{#1}          \fi
\ifx \showarticletitle \undefined \def \showarticletitle #1{#1}   \fi
\ifx \showURL      \undefined \def \showURL       {\relax}        \fi
\providecommand\bibfield[2]{#2}
\providecommand\bibinfo[2]{#2}
\providecommand\natexlab[1]{#1}
\providecommand\showeprint[2][]{arXiv:#2}

\bibitem[\protect\citeauthoryear{Aleahmad, Karisani, Rahgozar, and
  Oroumchian}{Aleahmad et~al\mbox{.}}{2015}]%
        {aleahmad2015olfinder}
\bibfield{author}{\bibinfo{person}{Abolfazl Aleahmad}, \bibinfo{person}{Payam
  Karisani}, \bibinfo{person}{Maseud Rahgozar}, {and} \bibinfo{person}{Farhad
  Oroumchian}.} \bibinfo{year}{2015}\natexlab{}.
\newblock \showarticletitle{OLFinder: Finding Opinion Leaders in Online Social
  Networks}.
\newblock \bibinfo{journal}{{\em Journal of Information Science\/}}
  \bibinfo{volume}{42}, \bibinfo{number}{5} (\bibinfo{date}{Sept.}
  \bibinfo{year}{2015}), \bibinfo{pages}{659--674}.
\newblock


\bibitem[\protect\citeauthoryear{Allport}{Allport}{1979}]%
        {allport1979nature}
\bibfield{author}{\bibinfo{person}{Gordon~W. Allport}.}
  \bibinfo{year}{1979}\natexlab{}.
\newblock \bibinfo{booktitle}{{\em The Nature of Prejudice: 25th Anniversary
  Edition}}.
\newblock \bibinfo{publisher}{Basic Books}, \bibinfo{address}{New York, NY,
  USA}, Chapter Patterning and Extent of Prejudice, \bibinfo{pages}{68--81}.
\newblock


\bibitem[\protect\citeauthoryear{Anderson-Clark, Green, and
  Henley}{Anderson-Clark et~al\mbox{.}}{2008}]%
        {anderson-clark2008relationship}
\bibfield{author}{\bibinfo{person}{Tracy~N. Anderson-Clark},
  \bibinfo{person}{Raymond~J. Green}, {and} \bibinfo{person}{Tracy~B. Henley}.}
  \bibinfo{year}{2008}\natexlab{}.
\newblock \showarticletitle{The Relationship Between First Names and Teacher
  Expectations for Achievement Motivation}.
\newblock \bibinfo{journal}{{\em Journal of Language and Social Psychology\/}}
  \bibinfo{volume}{27}, \bibinfo{number}{1} (\bibinfo{date}{March}
  \bibinfo{year}{2008}), \bibinfo{pages}{94--99}.
\newblock


\bibitem[\protect\citeauthoryear{Bandari, Asur, and Huberman}{Bandari
  et~al\mbox{.}}{2012}]%
        {bandari2012pulse}
\bibfield{author}{\bibinfo{person}{Roja Bandari}, \bibinfo{person}{Sitaram
  Asur}, {and} \bibinfo{person}{Bernardo Huberman}.}
  \bibinfo{year}{2012}\natexlab{}.
\newblock \showarticletitle{The Pulse of News in Social Media: Forecasting
  Popularity}. In \bibinfo{booktitle}{{\em 6th International Conference on Web
  and Social Media (ICWSM-12), Proceedings}}. \bibinfo{publisher}{AAAI},
  \bibinfo{address}{Palo Alto, CA, USA}, \bibinfo{pages}{26--33}.
\newblock


\bibitem[\protect\citeauthoryear{Bertrand and Mullainathan}{Bertrand and
  Mullainathan}{2003}]%
        {bertrand2003are}
\bibfield{author}{\bibinfo{person}{Marianne Bertrand} {and}
  \bibinfo{person}{Sendhil Mullainathan}.} \bibinfo{year}{2003}\natexlab{}.
\newblock \bibinfo{booktitle}{{\em Are Emily and Greg More Employable than
  Lakisha and Jamal? A Field Experiment on Labor Market Discrimination}}.
\newblock \bibinfo{type}{Working Paper} 9873. \bibinfo{institution}{National
  Bureau of Economic Research}.
\newblock


\bibitem[\protect\citeauthoryear{Bigonha, Cardoso, Moro, Gonçalves, and
  Almeida}{Bigonha et~al\mbox{.}}{2012}]%
        {bigonha2012sentiment}
\bibfield{author}{\bibinfo{person}{Carolina Bigonha}, \bibinfo{person}{Thiago
  N.~C. Cardoso}, \bibinfo{person}{Mirella~M. Moro}, \bibinfo{person}{Marcos~A.
  Gonçalves}, {and} \bibinfo{person}{Virgílio A.~F. Almeida}.}
  \bibinfo{year}{2012}\natexlab{}.
\newblock \showarticletitle{Sentiment-based Influence Detection on Twitter}.
\newblock \bibinfo{journal}{{\em Journal of the Brazilian Computer Society\/}}
  \bibinfo{volume}{18}, \bibinfo{number}{3} (\bibinfo{date}{Sept.}
  \bibinfo{year}{2012}), \bibinfo{pages}{169--183}.
\newblock


\bibitem[\protect\citeauthoryear{Cappelletti and Sastry}{Cappelletti and
  Sastry}{2012}]%
        {cappelletti2012iarank}
\bibfield{author}{\bibinfo{person}{Rafael Cappelletti} {and}
  \bibinfo{person}{Nishanth Sastry}.} \bibinfo{year}{2012}\natexlab{}.
\newblock \showarticletitle{IARank: Ranking Users on Twitter in Near Real-Time,
  Based on Their Information Amplification Potential}. In
  \bibinfo{booktitle}{{\em 1st International Conference on Social Informatics
  (SocialInformatics 2012), Proceedings}}. \bibinfo{publisher}{IEEE},
  \bibinfo{address}{New York, NY, USA}, \bibinfo{pages}{70--77}.
\newblock


\bibitem[\protect\citeauthoryear{Chorley, Colombo, Allen, and Whitaker}{Chorley
  et~al\mbox{.}}{2015}]%
        {chorley2015human}
\bibfield{author}{\bibinfo{person}{Martin~J. Chorley},
  \bibinfo{person}{Gualtiero~B. Colombo}, \bibinfo{person}{Stuart~M. Allen},
  {and} \bibinfo{person}{Roger~M. Whitaker}.} \bibinfo{year}{2015}\natexlab{}.
\newblock \showarticletitle{Human Content Filtering in Twitter: The Influence
  of Metadata}.
\newblock \bibinfo{journal}{{\em International Journal of Human-Computer
  Studies\/}}  \bibinfo{volume}{74} (\bibinfo{date}{Feb.}
  \bibinfo{year}{2015}), \bibinfo{pages}{32--40}.
\newblock


\bibitem[\protect\citeauthoryear{Coffey and McLaughlin}{Coffey and
  McLaughlin}{2009}]%
        {coffey2009do}
\bibfield{author}{\bibinfo{person}{Bentley Coffey} {and}
  \bibinfo{person}{Patrick~A. McLaughlin}.} \bibinfo{year}{2009}\natexlab{}.
\newblock \showarticletitle{Do Masculine Names Help Female Lawyers Become
  Judges? Evidence from South Carolina}.
\newblock \bibinfo{journal}{{\em American Law and Economics Review\/}}
  \bibinfo{volume}{11}, \bibinfo{number}{1} (\bibinfo{date}{Aug.}
  \bibinfo{year}{2009}), \bibinfo{pages}{112--133}.
\newblock


\bibitem[\protect\citeauthoryear{Fryer and Levitt}{Fryer and Levitt}{2004}]%
        {fryer2004causes}
\bibfield{author}{\bibinfo{person}{Roland~G. Fryer} {and}
  \bibinfo{person}{Steven~D. Levitt}.} \bibinfo{year}{2004}\natexlab{}.
\newblock \showarticletitle{The Causes and Consequences of Distinctively Black
  Names}.
\newblock \bibinfo{journal}{{\em The Quarterly Journal of Economics\/}}
  \bibinfo{volume}{119}, \bibinfo{number}{3} (\bibinfo{date}{Aug.}
  \bibinfo{year}{2004}), \bibinfo{pages}{767--805}.
\newblock


\bibitem[\protect\citeauthoryear{Gaudeul and Giannetti}{Gaudeul and
  Giannetti}{2015}]%
        {gaudeul2015privacy}
\bibfield{author}{\bibinfo{person}{Alexia Gaudeul} {and}
  \bibinfo{person}{Caterina Giannetti}.} \bibinfo{year}{2015}\natexlab{}.
\newblock \bibinfo{booktitle}{{\em Privacy, Trust and Social Network
  Formation}}.
\newblock \bibinfo{type}{Jena Economic Research Papers} 2015-023.
  \bibinfo{institution}{Friedrich-Schiller-University Jena}.
\newblock


\bibitem[\protect\citeauthoryear{Gupta, Goel, Lin, Sharma, Wang, and
  Zadeh}{Gupta et~al\mbox{.}}{2013}]%
        {gupta2013wtf}
\bibfield{author}{\bibinfo{person}{Pankaj Gupta}, \bibinfo{person}{Ashish
  Goel}, \bibinfo{person}{Jimmy Lin}, \bibinfo{person}{Aneesh Sharma},
  \bibinfo{person}{Dong Wang}, {and} \bibinfo{person}{Reza Zadeh}.}
  \bibinfo{year}{2013}\natexlab{}.
\newblock \showarticletitle{{WTF}: The Who to Follow Service at Twitter}. In
  \bibinfo{booktitle}{{\em 22nd International Conference on World Wide Web (WWW
  2013), Proceedings}}. \bibinfo{publisher}{ACM}, \bibinfo{address}{New York,
  NY, USA}, \bibinfo{pages}{505--514}.
\newblock


\bibitem[\protect\citeauthoryear{Hannon, Bennett, and Smyth}{Hannon
  et~al\mbox{.}}{2010}]%
        {hannon2010recommending}
\bibfield{author}{\bibinfo{person}{John Hannon}, \bibinfo{person}{Mike
  Bennett}, {and} \bibinfo{person}{Barry Smyth}.}
  \bibinfo{year}{2010}\natexlab{}.
\newblock \showarticletitle{Recommending Twitter Users to Follow Using Content
  and Collaborative Filtering Approaches}. In \bibinfo{booktitle}{{\em 4th
  Conference on Recommender Systems (RecSys 2010), Proceedings}}.
  \bibinfo{publisher}{ACM}, \bibinfo{address}{New York, NY, USA},
  \bibinfo{pages}{199--206}.
\newblock


\bibitem[\protect\citeauthoryear{Hong, Dan, and Davison}{Hong
  et~al\mbox{.}}{2011}]%
        {hong2011predicting}
\bibfield{author}{\bibinfo{person}{Liangjie Hong}, \bibinfo{person}{Ovidiu
  Dan}, {and} \bibinfo{person}{Brian~D. Davison}.}
  \bibinfo{year}{2011}\natexlab{}.
\newblock \showarticletitle{Predicting Popular Messages in Twitter}. In
  \bibinfo{booktitle}{{\em 20th International Conference Companion on World
  Wide Web (WWW 2011), Proceedings}}. \bibinfo{publisher}{ACM},
  \bibinfo{address}{New York, NY, USA}, \bibinfo{pages}{57--58}.
\newblock
\newblock
\shownote{Poster.}


\bibitem[\protect\citeauthoryear{Hutto, Yardi, and Gilbert}{Hutto
  et~al\mbox{.}}{2013}]%
        {hutto2013longitudinal}
\bibfield{author}{\bibinfo{person}{Clayton~J. Hutto}, \bibinfo{person}{Sarita
  Yardi}, {and} \bibinfo{person}{Eric Gilbert}.}
  \bibinfo{year}{2013}\natexlab{}.
\newblock \showarticletitle{A Longitudinal Study of Follow Predictors on
  Twitter}. In \bibinfo{booktitle}{{\em 32nd Conference on Human Factors in
  Computing Systems (CHI 2013), Proceedings}}. \bibinfo{publisher}{ACM},
  \bibinfo{address}{New York, NY, USA}, \bibinfo{pages}{821--830}.
\newblock


\bibitem[\protect\citeauthoryear{Kaiser}{Kaiser}{2009}]%
        {kaiser2009vornamen}
\bibfield{author}{\bibinfo{person}{Astrid Kaiser}.}
  \bibinfo{year}{2009}\natexlab{}.
\newblock \showarticletitle{Vornamen: Nomen est omen? [Given Names: Nomen est
  omen?]}.
\newblock \bibinfo{journal}{{\em Oberfränkischer Schulanzeiger\/}}
  \bibinfo{number}{12} (\bibinfo{date}{Dec.} \bibinfo{year}{2009}),
  \bibinfo{pages}{15--18}.
\newblock


\bibitem[\protect\citeauthoryear{Kwak, Lee, Park, and Moon}{Kwak
  et~al\mbox{.}}{2010}]%
        {kwak2010what}
\bibfield{author}{\bibinfo{person}{Haewoon Kwak}, \bibinfo{person}{Changhyun
  Lee}, \bibinfo{person}{Hosung Park}, {and} \bibinfo{person}{Sue Moon}.}
  \bibinfo{year}{2010}\natexlab{}.
\newblock \showarticletitle{What is Twitter, a Social Network or a News
  Media?}. In \bibinfo{booktitle}{{\em 19th International Conference on World
  Wide Web (WWW 2010), Proceedings}}. \bibinfo{publisher}{ACM},
  \bibinfo{address}{New York, NY, USA}, \bibinfo{pages}{591--600}.
\newblock


\bibitem[\protect\citeauthoryear{Martín, Lavesson, and Doroud}{Martín
  et~al\mbox{.}}{2016}]%
        {martin2016hashtags}
\bibfield{author}{\bibinfo{person}{Eva~García Martín},
  \bibinfo{person}{Niklas Lavesson}, {and} \bibinfo{person}{Mina Doroud}.}
  \bibinfo{year}{2016}\natexlab{}.
\newblock \showarticletitle{Hashtags and Followers}.
\newblock \bibinfo{journal}{{\em Social Network Analysis and Mining\/}}
  \bibinfo{volume}{6}, \bibinfo{number}{1} (\bibinfo{date}{March}
  \bibinfo{year}{2016}), \bibinfo{pages}{12}.
\newblock


\bibitem[\protect\citeauthoryear{McPherson, Smith-Lovin, and Cook}{McPherson
  et~al\mbox{.}}{2001}]%
        {mcpherson2001birds}
\bibfield{author}{\bibinfo{person}{Miller McPherson}, \bibinfo{person}{Lynn
  Smith-Lovin}, {and} \bibinfo{person}{James~M. Cook}.}
  \bibinfo{year}{2001}\natexlab{}.
\newblock \showarticletitle{Birds of a Feather: Homophily in Social Networks}.
\newblock \bibinfo{journal}{{\em Annual Review of Sociology\/}}
  \bibinfo{volume}{27} (\bibinfo{date}{Aug.} \bibinfo{year}{2001}),
  \bibinfo{pages}{415--444}.
\newblock


\bibitem[\protect\citeauthoryear{Nagmoti, Teredesai, and De~Cock}{Nagmoti
  et~al\mbox{.}}{2010}]%
        {nagmoti2010ranking}
\bibfield{author}{\bibinfo{person}{Rinkesh Nagmoti}, \bibinfo{person}{Ankur
  Teredesai}, {and} \bibinfo{person}{Martine De~Cock}.}
  \bibinfo{year}{2010}\natexlab{}.
\newblock \showarticletitle{Ranking Approaches for Microblog Search}. In
  \bibinfo{booktitle}{{\em International Conference on Web Intelligence and
  Intelligent Agent Technology (WI-IAT 2010), Proceedings}},
  Vol.~\bibinfo{volume}{1}. \bibinfo{publisher}{IEEE}, \bibinfo{address}{New
  York, NY, USA}, \bibinfo{pages}{153--157}.
\newblock


\bibitem[\protect\citeauthoryear{Noro, Ru, Xiao, and Tokuda}{Noro
  et~al\mbox{.}}{2013}]%
        {noro2013twitter}
\bibfield{author}{\bibinfo{person}{Tomoya Noro}, \bibinfo{person}{Fei Ru},
  \bibinfo{person}{Feng Xiao}, {and} \bibinfo{person}{Takehiro Tokuda}.}
  \bibinfo{year}{2013}\natexlab{}.
\newblock \showarticletitle{Twitter User Rank Using Keyword Search}. In
  \bibinfo{booktitle}{{\em Information Modelling and Knowledge Bases XXIV}}
  {\em (\bibinfo{series}{Frontiers in Artificial Intelligence and
  Applications})}, Vol.~\bibinfo{volume}{251}. \bibinfo{publisher}{IOS Press},
  \bibinfo{address}{Amsterdam, Netherlands}, \bibinfo{pages}{31--48}.
\newblock


\bibitem[\protect\citeauthoryear{Oliver, Wood, and Bass}{Oliver
  et~al\mbox{.}}{2015}]%
        {oliver2015liberellas}
\bibfield{author}{\bibinfo{person}{J.~Eric Oliver}, \bibinfo{person}{Thomas
  Wood}, {and} \bibinfo{person}{Alexandra Bass}.}
  \bibinfo{year}{2015}\natexlab{}.
\newblock \showarticletitle{Liberellas Versus Konservatives: Social Status,
  Ideology, and Birth Names in the United States}.
\newblock \bibinfo{journal}{{\em Political Behavior\/}} \bibinfo{volume}{38},
  \bibinfo{number}{1} (\bibinfo{date}{March} \bibinfo{year}{2015}),
  \bibinfo{pages}{55--81}.
\newblock


\bibitem[\protect\citeauthoryear{Razis and Anagnostopoulos}{Razis and
  Anagnostopoulos}{2014}]%
        {razis2014influencetracker}
\bibfield{author}{\bibinfo{person}{Gerasimos Razis} {and}
  \bibinfo{person}{Ioannis Anagnostopoulos}.} \bibinfo{year}{2014}\natexlab{}.
\newblock \showarticletitle{InfluenceTracker: Rating the Impact of a Twitter
  Account}.
\newblock In \bibinfo{booktitle}{{\em Artificial Intelligence Applications and
  Innovations}}. \bibinfo{series}{IFIP Advances in Information and
  Communication Technology}, Vol.~\bibinfo{volume}{437}.
  \bibinfo{publisher}{Springer}, \bibinfo{address}{Berlin / Heidelberg,
  Germany}, \bibinfo{pages}{184--195}.
\newblock


\bibitem[\protect\citeauthoryear{Riquelme and González-Cantergiani}{Riquelme
  and González-Cantergiani}{2016}]%
        {riquelme2016measuring}
\bibfield{author}{\bibinfo{person}{Fabián Riquelme} {and}
  \bibinfo{person}{Pablo González-Cantergiani}.}
  \bibinfo{year}{2016}\natexlab{}.
\newblock \showarticletitle{Measuring User Influence on Twitter: A Survey}.
\newblock \bibinfo{journal}{{\em Information Processing \& Management\/}}
  \bibinfo{volume}{52}, \bibinfo{number}{5} (\bibinfo{date}{Sept.}
  \bibinfo{year}{2016}), \bibinfo{pages}{949--975}.
\newblock


\bibitem[\protect\citeauthoryear{Srinivasan, Srinivasa, and
  Thulasidasan}{Srinivasan et~al\mbox{.}}{2014}]%
        {srinivasan2014comparative}
\bibfield{author}{\bibinfo{person}{MS Srinivasan}, \bibinfo{person}{Srinath
  Srinivasa}, {and} \bibinfo{person}{Sunil Thulasidasan}.}
  \bibinfo{year}{2014}\natexlab{}.
\newblock \showarticletitle{A Comparative Study of Two Models for Celebrity
  Identification on Twitter}. In \bibinfo{booktitle}{{\em 20th International
  Conference on Management of Data (COMAD 2014), Proceedings}}.
  \bibinfo{publisher}{Computer Society of India}, \bibinfo{address}{Mumbai,
  India, India}, \bibinfo{pages}{57--65}.
\newblock


\bibitem[\protect\citeauthoryear{Suh, Hong, Pirolli, and Chi}{Suh
  et~al\mbox{.}}{2010}]%
        {suh2010want}
\bibfield{author}{\bibinfo{person}{Bongwon Suh}, \bibinfo{person}{Lichan Hong},
  \bibinfo{person}{Peter Pirolli}, {and} \bibinfo{person}{Ed~H. Chi}.}
  \bibinfo{year}{2010}\natexlab{}.
\newblock \showarticletitle{Want to be Retweeted? Large Scale Analytics on
  Factors Impacting Retweet in Twitter Network}. In \bibinfo{booktitle}{{\em
  2nd International Conference on Social Computing (SocialCom 2010),
  Proceedings}}. \bibinfo{publisher}{IEEE}, \bibinfo{address}{New York, NY,
  USA}, \bibinfo{pages}{177--184}.
\newblock


\bibitem[\protect\citeauthoryear{Thorndike}{Thorndike}{1920}]%
        {thorndike1920constant}
\bibfield{author}{\bibinfo{person}{Edward~L. Thorndike}.}
  \bibinfo{year}{1920}\natexlab{}.
\newblock \showarticletitle{A Constant Error in Psychological Ratings}.
\newblock \bibinfo{journal}{{\em Journal of Applied Psychology\/}}
  \bibinfo{volume}{4}, \bibinfo{number}{1} (\bibinfo{date}{March}
  \bibinfo{year}{1920}), \bibinfo{pages}{25--29}.
\newblock


\bibitem[\protect\citeauthoryear{Tsur and Rappoport}{Tsur and
  Rappoport}{2015}]%
        {tsur2015dont}
\bibfield{author}{\bibinfo{person}{Oren Tsur} {and} \bibinfo{person}{Ari
  Rappoport}.} \bibinfo{year}{2015}\natexlab{}.
\newblock \showarticletitle{Don’t Let Me Be \#Misunderstood: Linguistically
  Motivated Algorithm for Predicting the Popularity of Textual Memes}. In
  \bibinfo{booktitle}{{\em 9th International Conference on Web and Social Media
  (ICWSM-15), Proceedings}}. \bibinfo{publisher}{AAAI}, \bibinfo{address}{Palo
  Alto, CA, USA}, \bibinfo{pages}{426--435}.
\newblock


\bibitem[\protect\citeauthoryear{{Twitter, Inc.}}{{Twitter, Inc.}}{2015}]%
        {twitter2015annual}
\bibfield{author}{\bibinfo{person}{{Twitter, Inc.}}}
  \bibinfo{year}{2015}\natexlab{}.
\newblock \bibinfo{title}{Annual Report 2016}.
\newblock San Francisco, CA, USA.
\newblock


\bibitem[\protect\citeauthoryear{Weng, Lim, Jiang, and He}{Weng
  et~al\mbox{.}}{2010}]%
        {weng2010twitterrank}
\bibfield{author}{\bibinfo{person}{Jianshu Weng}, \bibinfo{person}{Ee-Peng
  Lim}, \bibinfo{person}{Jing Jiang}, {and} \bibinfo{person}{Qi He}.}
  \bibinfo{year}{2010}\natexlab{}.
\newblock \showarticletitle{TwitterRank: Finding Topic-sensitive Influential
  Twitterers}. In \bibinfo{booktitle}{{\em 3rd International Conference on Web
  Search and Data Mining (WSDM 2010), Proceedings}}. \bibinfo{publisher}{ACM},
  \bibinfo{address}{New York, NY, USA}, \bibinfo{pages}{261--270}.
\newblock


\end{thebibliography}

\end{document}